# Half Heusler alloy CoVSn as self-supported electrocatalyst for hydrogen evolution reaction

*Deepak Gujjar[1] and Hem C. Kandpal[*1]*

[1]Department of Chemistry, Indian Institute of Technology Roorkee, Roorkee, Uttarakhand - 247667, India

*Email address: hem.kandpal@cy.iitr.ac.in , Phone: +91-1332-28-4764.

**Abstract**

Despite significant advancements in electrocatalysis for clean hydrogen fuel generation, the transition from concept to commercialization faces challenges due to the instability of electrocatalysts. This study delves into the exploration of a structurally and mechanically robust half-Heusler alloy, CoVSn, as an efficient electrocatalyst for hydrogen production. The synthesis of CoVSn was achieved using the arc-melting technique and optimized successfully into a cubic structure – a previously unattained and highly challenging feat. The resulting electrode, cut from the obtained CoVSn pellet, served as a self-supported electrocatalyst and initially generates a current density of −10 mA cm$^{−2}$ at an overpotential of 244 mV. Remarkably, this overpotential decreased uniquely over time, reaches 202 mV after a durability testing of 12 hours, while maintaining its crystal structure integrity after the electrocatalysis process. This progressive enhancement in catalytic activity and structural stability underscores the significance of this research. The synergistic effect between Co and V atoms as pivotal active centres for hydrogen generation was evident, further enhanced by formation of high valance metal sites $Co_2O_3$ and $V_2O_3$ during the hydrogen evolution reaction. In essence, this study confirms the stability and promise of CoVSn in hydrogen generation, paving the way for exploring additional self-supported ternary intermetallics to enhance water-splitting efficiency.

## 1. Introduction

Intermetallic compounds, a prominent class of materials, possess diverse applications across various fields, including thermoelectricity, superconductivity, shape memory effects, and spintronics.[1] Their remarkable crystallographic and electronic properties have sparked interest among researchers, leading to explorations in the realm of heterogeneous catalysis.[2] Particularly, there has been a recent surge in investigating their potential as electrocatalysts in electrochemical water splitting to generate clean hydrogen fuel, aiming to offer a cost-effective alternative to the conventional, expensive Pt-based electrocatalysts. Several non-noble binary intermetallics like $CoSn_2$,[3] NiZn,[4] $NiAl_3$,[5] $MoNi_4$,[6] $Co_3Mo$,[7] NiFe,[8] NiMo,[9] $Ni_2Si$,[10] $FeSb_2$[11] etc., have been used as electrocatalysts for the hydrogen evolution reaction (HER). However, studies by Lado et al. showcased improved electrocatalytic activity of AlNiP compared to its binary counterpart NiP,[12] while Pei et al. demonstrated superior performance of the ternary CoMoMg over its binary constituents CoMg and MoMg for hydrogen production.[13] This enhanced performance of ternary intermetallics is believed to arise from a higher number of interfaces between different types of atoms in their structure.[12] However, significant challenges endure, particularly concerning stability issues arising from element leaching within the electrolyte during the hydrogen evolution reaction (HER), along with the requirement for additional support to enable their fabrication as electrodes.

Our interest stems from a recent study where ternary Co-Mn-Sn alloys exhibited efficient electrocatalytic properties for HER, surpassing their single metal constituents (Co, Mn, Sn) and binary alloys (Co-Sn, Co-Mn).[14] However, this study involved electrodeposition of elements onto a copper support, and degradation was observed after 11 hours of durability testing, requiring a higher overpotential of 157 mV for a current density of -10 mA cm$^{-2}$.[14] Consequently, our pursuit focused on two objectives: identifying highly stable Co-Sn-based ternary intermetallics and optimizing a synthetic route for their direct electrode fabrication. The concept emerged from theoretical investigations in our group, where Zeeshan et al. highlighted the structural and mechanical robustness of half-Heusler CoVSn in its cubic phase, based on density functional theory (DFT) calculations.[15] Additionally, insights from literature, particularly Chen et al.'s work on the highly efficient electrocatalyst Ir$_3$V for hydrogen production, inspired the incorporation of vanadium into the system. DFT calculations in their study revealed that the improved HER activity of Ir$_3$V catalysts primarily stemmed from the enhanced water dissociation ability at the V sites.[16] It is essential to note that synthesizing pure c-CoVSn is a challenging endeavor due to the narrow range within the ternary phase diagram of Co-V-Sn, as reported by Shi et al.[17] Previous attempts by Zererani et al.[18] and Lue et al.[19] utilized spark plasma sintering (SPS) and RF induction furnace methods, respectively, to produce CoVSn. However, both studies observed the presence of CoVSn alongside various binary phases such as Co$_2$Sn, Co$_3$V, SnV$_3$, CoSn, and V$_2$Sn$_3$.

In our investigation, we successfully optimized half-Heusler CoVSn in its cubic phase using the arc-melting technique followed by vacuum-sealed sintering. From the synthesized ingot, an electrode of c-CoVSn was precisely cut using a diamond saw and employed directly as a self-supported electrode for the electrochemical generation of hydrogen.

## 2. Experimental Section

### 2.1. Chemicals

Co slug (3.175 mm dia, Alfa Aeser, 99.95 %), V slug (3.175 mm dia, Alfa Aeser, 99.8 %), and Sn slug (6.35 mm dia, Alfa Aeser, 99.995 %), were employed as metal precursors. The $H_2SO_4$ (Finar Chemicals, 98 %) were utilized for the preparation of electrolyte solutions for HER measurements. All the solutions were prepared in distilled water and all the metal ingots were used directly without any further purifications.

## 2.2. CoVSn Synthesis and Electrode Fabrication

The synthesis of half-Heusler alloy CoVSn was accomplished via the arc-melting technique. Stoichiometric amounts of metal ingots were meticulously weighed and introduced into the Cu-pit of an arc-melting chamber and the chamber's pressure was maintained at $10^{-6}$ mbar to eliminate atmospheric oxygen. To ensure the removal of any residual gases, argon gas was purged through the chamber thrice, and subsequently, the ingots underwent melting under high-temperature arc in an argon atmosphere, resulting in the formation of a CoVSn pellet. This melting process was iterated six times to ensure the uniform diffusion of metals within the pellet. The resulting CoVSn pellet enclosed within a vacuum-sealed quartz tube underwent further temperature treatment for an additional 15 days in a muffle furnace. This prolonged temperature treatment was employed to achieve complete homogenization and optimize the desired phase. Three separate samples were synthesized with varied temperature treatments of 900, 850, and 700 °C. Notably, the most desirable sample, exhibiting minimal impurity phase of Sn, was obtained at 700°C.

The CoVSn pellet obtained from the optimized process was sectioned using a diamond saw to produce a CoVSn bar, which served as the working electrode for subsequent electrocatalytic measurements.

## 3. Result and discussion

## 3.1. Structural and Morphological Analysis

Fig. 1 displays the X-ray diffraction (XRD) profiles of CoVSn samples fabricated via arc-melting, employing diverse temperature treatments. All the patterns exhibit analogous characteristics, signifying the crystallization of CoVSn into a cubic structure with a *P-43m* space group, aligning well with the XRD pattern obtained after the optimized lattice parameters as previously reported by Zeeshan et al.[15] Predominant peaks observed correspond to the crystallographic planes (111), (200), (220), (311), (400), and (422). Notably, all samples exhibit the presence of Sn phase impurities; however, optimization procedures have identified the 700 °C treatment as yielding the most favorable sample. It is noteworthy that the attainment of a pure CoVSn phase remains elusive, as prior investigations have reported the synthesis of CoVSn alongside diverse binary phases such as $Co_2Sn$, $Co_3V$, $SnV_3$, CoSn, and $V_2Sn_3$.[18,19] However, our methodology successfully mitigates these binary phases, presenting optimized conditions conducive to minimizing their presence.

The morphological analysis of CoVSn pellet were performed using field emission scanning electron microscopy (FESEM) coupled with energy dispersive X-ray analysis (EDXA) to precisely ascertain its elemental composition. The FESEM image (Fig. 2a) clearly shows the smooth and defect-free surface of the sample, yet discernible small white circular patches were also observed. Elemental mapping (Fig. 2b-2e) unequivocally indicates the dominance of Sn within these white patches, aligning closely with the X-ray diffraction findings suggesting the presence of additional Sn phases in the sample. Despite this, the EDS analysis shown in Fig. 2f, validates the overall sample composition in the desired stoichiometry Additionally, confirmation of CoVSn formation was reinforced via X-ray photoelectron spectroscopy (XPS), as elaborated in section 3.3 (post HER characterizations).

## 3.2. Electrocatalytic measurements

Following the successful characterization, the CoVSn bar was directly utilized as the working electrode for the hydrogen evolution reaction (HER) in a standard three-electrode system, employing a Pt wire as the counter electrode and Ag/AgCl/Cl⁻ as the reference electrode. The experiments were conducted at room temperature using a 0.05 M $H_2SO_4$ electrolytic solution. The measurements were performed at a scan rate of 10 mV s$^{-1}$ within the potential range of -0.5 to 0.5 V vs reversible hydrogen electrode (RHE), utilizing a CoVSn bar with an immersed geometric surface area of 0.10 cm$^2$.

The initial linear sweep voltammetry (LSV) polarization curve was recorded following the activation of the CoVSn bar through 100 stabilized cyclic voltammetry (CV) cycles. The outcomes (Fig. 3a) demonstrated commendable HER catalytic activity, where the CoVSn bar exhibited a current density of −10 mA cm$^{-2}$ with an IR-corrected overpotential ($\eta$) value of 244 mV. To evaluate the durability of the CoVSn electrode, a chronoamperometric (CA) measurement was conducted at a constant overpotential of −0.2 V vs RHE for 12 hours. The results (Fig. 3b) exhibited consistent current density generation, indicating its potential practical and industrial viability. Notably, a progressive enhancement in current density over time (Fig. 3b, inset) was observed, signifies an augmented electrocatalytic activity of the electrode. However, an observation of slight decrease in current density around the 5-hours mark is likely attributed to the addition of electrolyte at that moment. The polarization curve obtained after the 12-hour CA measurements reaffirmed the increased activity, requiring an overpotential (IR corrected) as low as 202 mV, 42 mV less than the initial value, to achieve a current density of −10 mA cm$^{-2}$. These results surpass the performance of previously reported arc-melted ternary alloys Co-Ni-R (R = Y, Pr, Er) electrocatalysts, which generated lower current densities of 9.3, 4.9, and 6.4 mA cm$^{-2}$ with an overpotential of 200 mV.[20]

Moreover, the corresponding Tafel slopes pre- and post-durability testing were evaluated to analyse the kinetics governing hydrogen evolution. The Tafel slopes were derived by fitting the IR corrected overpotential ($\eta$) values against the logarithm of current density (Fig. 3c). The decrease in the Tafel slope value from 163 to 101 mV dec$^{-1}$ after durability testing suggest faster electron exchange, affirming the activation of the CoVSn electrode. The observed Tafel slope of 101 mV dec$^{-1}$ predominantly indicates the rate-determining Volmer step ($H_3O^+ + e^- + M \rightleftharpoons M-H + H_2O$), where hydrogen adsorption occurs on vacant sites (M) across the electrocatalyst surface. Subsequent hydrogen generation takes place in the second step, either through the Heyrovsky mechanism ($M-H + H_3O^+ + e^- \rightleftharpoons H_2 + H_2O + M$) or the Tafel step ($2M-H \rightleftharpoons H_2 + 2M$). The parallel observed Tafel slope value has been previously documented by Bonde et al. for ternary Co–Mo–S electrocatalyst.[21] Additionally, electrochemical impedance spectra (EIS) were obtained before and after CA measurements and depicted as a Nyquist plot in Fig. 3d. The acquired data points were fitted with an appropriate equivalent circuit (Fig. 3d, inset) to determine the uncompensated ($R_u$) and charge transfer ($R_{ct}$) resistances. The $R_{ct}$ value after 12 hours of durability testing reduced significantly from 2.427 × 10$^4$ to 1896 $\Omega$, reflecting improved catalytic performance. The reduction in semicircle diameters corroborated these findings. Nevertheless, it is noteworthy that the initial and post-CA measurement $R_u$ values were determined as 5.5 and 4.462 $\Omega$, respectively, signifies a relatively minor alteration in the uncompensated resistance throughout the experimental process.

Further analysis involved determining the double layer charge capacitance ($C_{dl}$) from cyclic voltammetry curves recorded across various scan rates (10 – 500 mV s$^{-1}$) in the non-Faradaic region (0.02 to 0.26 V vs RHE) to evaluate the electrochemically active surface area (ECSA) of the working electrode (Fig. 3e). The resulting $C_{dl}$ value of 0.596 mF cm$^{-2}$ (Fig. 3f) indicates efficient electron transport at the interfaces of the electrocatalyst and electrolyte. This value, nearly three times higher than that of single-walled carbon nanotubes (SWCNT),[22] underscores the efficacy of the self-supported electrode methodology in providing a greater number of electroactive sites for hydrogen generation, thereby enhancing electrocatalytic performance.

### 3.3. Post HER Characterizations & $H_2$ generation mechanism

Investigating the stability of the electrocatalyst requires careful evaluation of any structural, morphological, and compositional modifications after the electrochemical process. Therefore, we undertook an extensive assessment of the CoVSn bar subsequent to the hydrogen evolution reaction (HER) employing XRD, SEM, EDXA, and XPS techniques. The post-HER XRD pattern (supporting information: Fig. S1) prominently showcases the preserved crystal structure of c-CoVSn, dismissing the likelihood of leaching from the electrode. However, the emergence of some minor peaks indicates the formation of cobalt ($Co_2O_3$) and vanadium ($V_2O_3$) oxides. Subsequent SEM imaging unveiled textural alterations, evident in the transformation of initially smooth white patches to rough areas and the EDS analysis performed on these regions validated the formation of $Co_2O_3$ (supporting information: Fig. S2a-S2d). Furthermore, the comparison of pre- and post-HER core-level XPS spectra related to Co 2p, V 2p, Sn 3d, and O 1s (Fig. 4), further affirmed the presence of $Co_2O_3$ and $V_2O_3$ oxides, aligning with the XRD and SEM observations. The Co 2p spectra (Fig. 4a) obtained before HER exhibit prominent peaks at 777.44 eV (Co $2p_{3/2}$) and 792.4 eV (Co $2p_{1/2}$), associated with the $Co^0$ or Co-Sn/V bond,[23,24] noticeably shifted to 778.06 eV (Co $2p_{3/2}$) and 793.0 eV (Co $2p_{1/2}$), respectively, after the HER.[25] The observed shift (Fig. 4a, inset) towards higher binding energy strongly suggests the possible formation of cobalt oxide. This assertion gains further support from the emergence of two additional peaks post-HER at 781.45 eV (Co $2p_{3/2}$) and 797.13 eV (Co $2p_{1/2}$), which distinctly correspond to the formation of $Co_2O_3$.[26] Correspondingly, the appearance of new peaks at 515.75 eV (V $2p_{3/2}$) and 523.3 eV (V $2p_{1/2}$) in the V 2p spectra (Fig. 4b) corroborates the presence of $V_2O_3$.[27] Moreover, alterations in the intensity of peaks linked to $Sn^0$ and $Sn^{4+}$ in Sn 3d spectra (Fig. 4c), alongwith in M–O bonds in O1s spectra (Fig. 4d), substantiated the oxide formation.[28]

Based on the post-HER outcomes, we posit that the collaborative action of $Co^0$ and $V^0$ active sites contributes to hydrogen generation[3] and the formation of $Co_2O_3$ and $V_2O_3$ during the HER further promote its electrocatalytic activity.[29] The high valance $Co^{3+}$ exhibits an enhanced tendency to capture electrons from the cathode at applied potential, facilitated by good metallic conductivity. This aids in electron transfer to surrounding $H^+$ ions, expediting the HER.[29] Additionally, V sites potentially synergize with Co-sites, accelerating the dissociation of water around the catalyst surface into $H^+$ ions.[16] These ions capture electrons from the catalyst surface to produce $H_2$. This mechanism is consistent with the recent report where Li et al. proposed the $Co^{3+}$ as highly active metal centre for stable and efficient HER.[29] Importantly, utilizing the highly dense arc-melted electrode directly proves more advantageous compared to using a compacted bar of chemically synthesized ternary alloys, as the latter often faces issues of leaching and stability, as evidenced in our previously reported FeNiAs system.[30]

## 4. Conclusions

In conclusion, this work documented the successful optimization of ternary CoVSn into a cubic structure via arc-melting, devoid of previously reported binary impurity phases, enabling direct utilization of the resulting highly dense arc-melted bar as an efficient electrocatalyst for water-splitting hydrogen generation. The utilization of self-supported sample bar effectively addressed the limitations of conventional powdery electrocatalysts, notably eliminating the need for expensive binders and conductive additives, while also mitigating leaching and stability issues typically associated with electrodeposited electrocatalysts. The observed long-term enhancement in electrocatalytic performance, attributed to the synergistic interplay between Co and V active sites during the HER, highlights the significance of this work.

Looking ahead, this work holds promise for expediting the journey towards commercialization. Additionally, it lays the groundwork for exploring other self-supported ternary intermetallics in the realm of electrocatalytic studies.

**AUTHOR INFORMATION**

**Corresponding Author**

**Acknowledgements**

We acknowledge the support of Institute Instrumentation Centre (IIC), IIT Roorkee for providing characterization facilities. D. Gujjar acknowledges the senior research fellowship (SRF) granted by University Grants Commission (UGC), India.

**Author Contributions**

**D. Gujjar:** Methodology, Conceptualization, Investigation, Data Curation, Validation, Formal analysis, Visualization, Writing-Original Draft, **Hem C. Kandpal:** Conceptualization, Supervision, Project administration, Funding acquisition, Writing-Review & Editing.

**Data availability statement**

The datasets generated or analysed during the current study will be available from the corresponding author upon reasonable request.

**Conflicts of Interest**

The authors have no relevant financial or non-financial interests to disclose.

**Supporting Information**

Details of material characterizations and electrocatalytic measurements and additional XRD, SEM, and EDXA data for CoVSn electrocatalyst after the HER.

**Figure legends**

**Fig. 1.** XRD pattern of CoVSn samples synthesized at different temperatures along with simulated CoVSn XRD pattern[15].

**Fig. 2.** SEM micrograph of CoVSn sample prepared at 700 °C (a), along with elemental mapping (b-e), and corresponding EDS analysis (f).

**Fig. 3.** HER polarization curves with and without IR-corrections (a), durability testing of CoVSn electrode using chronoamperometric measurements at constant a potential of −0.2 V vs RHE, showing electrode activation overtime as inset (b), corresponding pre- and post-CA measurement Tafel plots of CoVSn electrode (c), Pre- and post-CA measurement Nyquist plots of CoVSn electrode with equivalent circuit as inset (d), CV curves recorded in non-Faradaic region across various scan rates (e), and determination of $C_{dl}$ value from ΔJ/2 vs scan rate plot (f).

**Fig. 4.** Core-level XPS spectra of Co 2p (a), V 2p (b), Sn 3d (c), and O 1s (d) before and after HER.

**Figure 1**

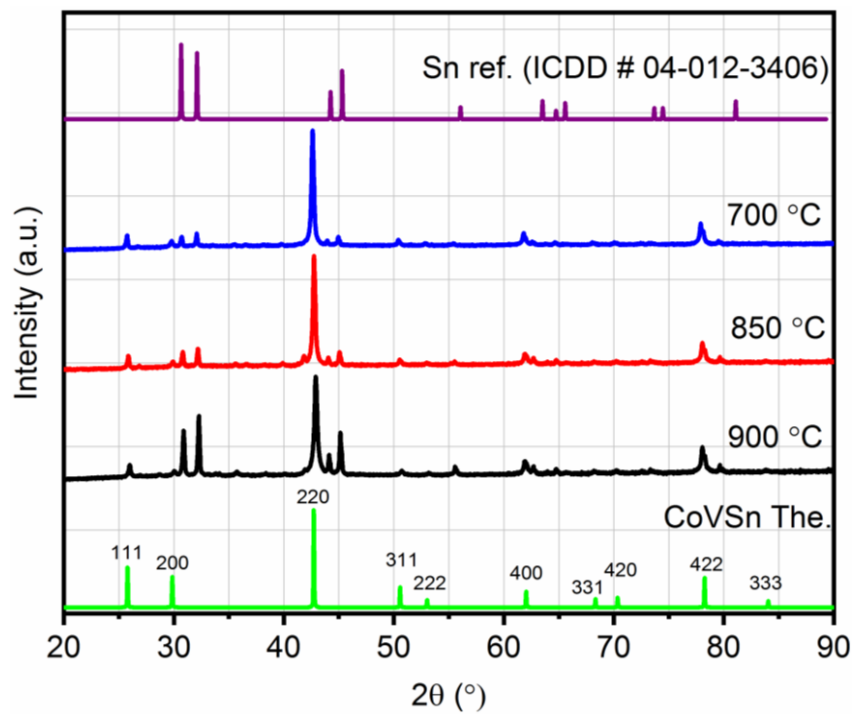

**Figure 2**

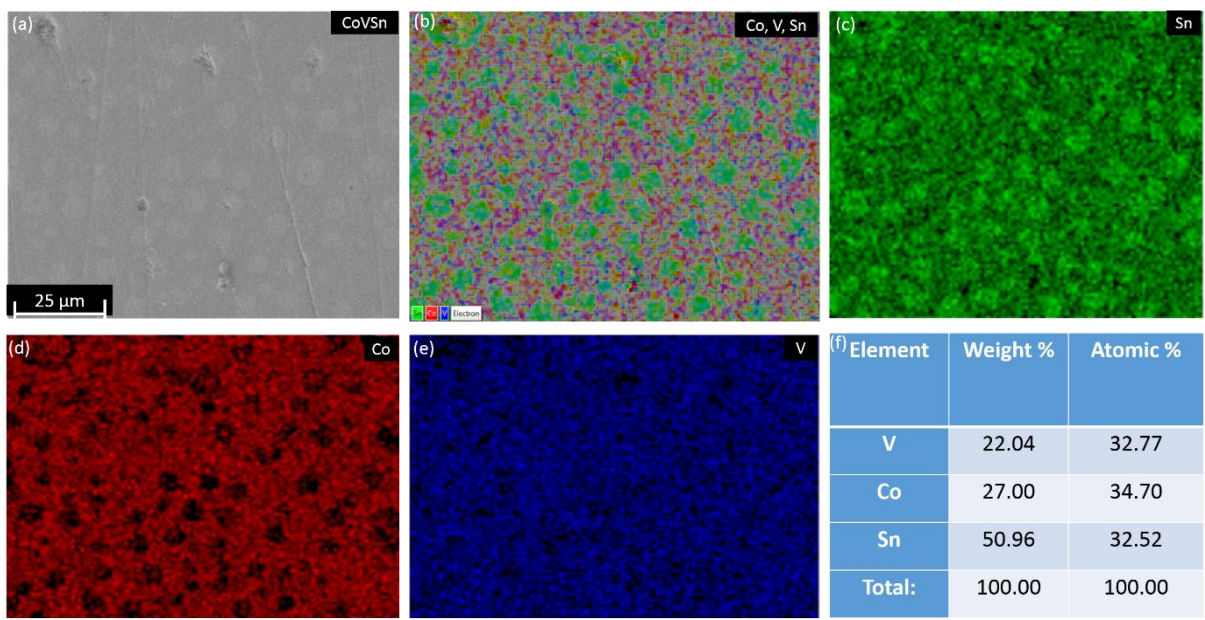

**Figure 3**

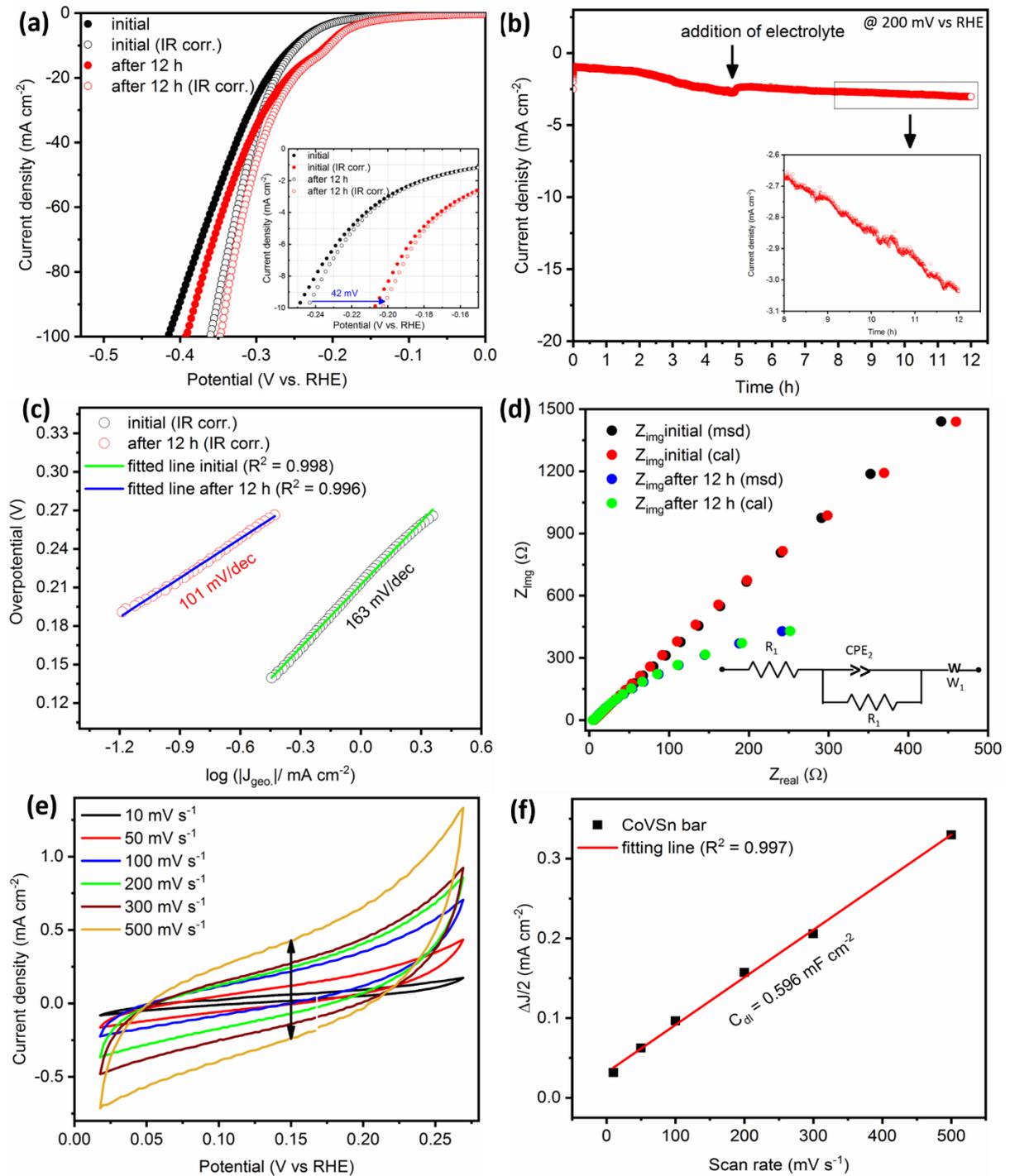

**Figure 4**

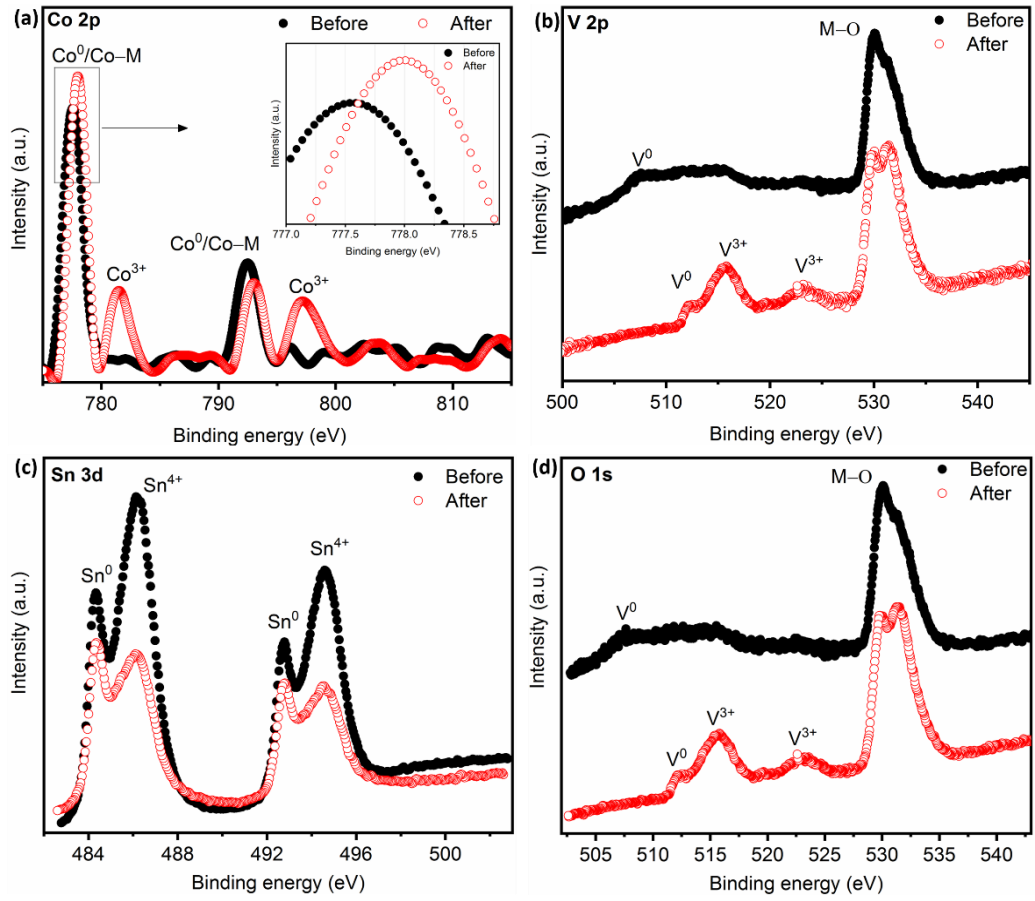